# Shared Understanding in Human–AI Teams: Coordination Continuity in Evolving Work

*Siân Lindley, Britta Burlin, Christian Poelitz, Lev Tankelevitch, and Richard Banks*

*Microsoft Research, Nov 2026*

## Abstract

People are increasingly working with AI agents as part of extended, asynchronous workflows. In these settings, shared access to information does not guarantee a shared basis for action: goals, plans, and roles may persist while their current significance changes. We introduce *coordination continuity*, the extent to which collaborators sustain a workable basis for interdependent action as work progresses. Drawing on research on collaborative work, along with two qualitative team studies and a reflexive case from our own repository-based environment, we present a framework of three elements: durable *representations* of collaborative work; relative *coordination properties* that support interpretation of representations in the moment; and *interaction mechanisms* through which the status of representations can be understood and altered. We use a *build* → *act* → *drift* → *rebuild* heuristic and the notion of *consequential moments* to characterise when rebuilding may be needed. We draw eight design implications for collaborative AI infrastructure.

**Keywords:** human–AI interaction; agentic AI; computer-supported cooperative work; asynchronous collaboration; articulation work; common ground; shared mental models; situation awareness; human–agent teams; handoff.

## Introduction

Human–AI interaction is moving beyond short, single-user sessions towards extended workflows. When people collaborate asynchronously, activity gets distributed across different people, artefacts, tools, and environments (Hutchins, 1995; Hollan *et al.*, 2000). Part of the work of working together is therefore building and sustaining a shared basis for action: collaborators need compatible understandings of what is being done, why it matters, and what should happen next. This is the role of *articulation work*: people selectively surface and negotiate aspects of the collaborative effort, so that everyone has the necessary context for their own part in the work (Schmidt & Bannon, 1992). Ultimately, collaborative outcomes depend on how well collaborators' understandings continue to hold together as the work progresses.

Introducing agents to collaborative work may change its distribution, but it does not change the need for the people and agents involved to develop congruent understandings[1] of the work in hand. Consider a scientific team in which one researcher determines the hypothesis, another oversees data-gathering, an agent writes analysis code, and a statistician reviews the result. Successful collaboration depends on more than providing each participant with information relevant to their immediate task. Participants need a workable understanding of how their actions knit together to address collective goals, who has expertise and availability, how decisions have been reached, and when something has changed enough to necessitate reconsideration.

---

[1] When we say that an agent "understands" some aspect of collaborative work, we use the term operationally to refer to the representations it maintains and the actions it produces in relation to team context, not to humanlike subjective comprehension. We therefore distinguish human mental models, machine representations, and observable compatibility of action.

In this paper, we build on the wealth of work in the field of CSCW (Computer-Supported Cooperative Work) to argue that providing shared context for human–agent activity is not in itself enough for collaborative work to succeed. Information must be continually related to situated practices and purposes (Bannon & Bødker, 1997), and understanding must be dynamically established (Clark & Brennan, 1991) and upheld on an ongoing basis (Schmidt & Bannon, 1992). This creates challenges in human–AI workflows where representations can persist in repositories or context windows, which are utilised by agents, whilst their relevance to the work changes. Contemporary infrastructures equip agents to retrieve repositories, follow specifications, act on plans, and generate work from stored artefacts. But coordinated work additionally requires collaborators to understand the current significance of the context provided through these resources. We refer to the extent to which a workable basis for interdependent action is sustained throughout a workflow as *coordination continuity*. Our focus is on how a workable basis for coordination can be sustained in human-agent teams, when artefacts and other representations of the work persist whilst the activity and meaning associated with them changes.

We ask three questions: *What* do people and agents need to understand together to act together? *How* can that understanding be built and maintained as work evolves? And, *when* should systems intervene to prevent or repair breakdowns in shared understanding?

The sections that follow develop our argument from theory through practice to design implications. We first draw on prior research to develop coordination continuity as a way of framing human–agent collaborative work. We then examine data from two qualitative team studies and a reflexive case from our own repository-based environment. We present a framework of three elements: durable *representations* of collaborative work; relative *coordination properties* that support interpretation of representations in the moment; and *interaction mechanisms* through which the status of representations can be understood and altered. We use a *build → act → drift → rebuild* heuristic and the notion of *consequential moments* to characterise when rebuilding may be needed. We finish with eight design implications for collaborative AI infrastructure.

# Related Work

## What must people and agents understand together to act together?

Shared Mental Model (SMM) theory proposes that teams coordinate more effectively when their members hold congruent understandings of the work, the environment, and each other. Cannon-Bowers *et al.* (1993) synthesised early work on SMMs, describing them as knowledge structures held in common by team members that cover aspects of the shared task (goals, procedures, equipment, constraints, and likely actions) and the team involved (roles, norms, expertise, and capabilities). Later work has often distinguished *taskwork* mental models from *teamwork* mental models, but the central point remains that teams need shared understanding both of what is being done and how to work together. Empirical work has repeatedly linked team cognition with coordination and performance (e.g., Mathieu *et al.*, 2000; DeChurch & Mesmer-Magnus, 2010).

Shared mental models do not require total overlap of knowledge. For instance, a research team's analysis may depend on a range of skills possessed by different experts in the group, alongside information held in code, notebooks, and visualisations. Coordination therefore depends on differentiated expertise (Hollingshead, 1998), knowledge of who knows what (Wegner, 1987; Lewis, 2003), and understanding of how artefacts, representations, and environments form part of the work (Hutchins, 1995). In extended human–AI workflows, the information and capabilities needed for collaboration are likely to be distributed across different people, agents, repositories, specifications, and tools. Participants may know different things, hold different interpretations, and may even disagree, but they must understand those differences well enough to coordinate.

Researchers have explored how to encode some aspects of shared mental models computationally. Scheutz *et al.* (2017), for example, make representations of goals, roles, capabilities, and beliefs explicit in support of human–robot interaction. Earlier multi-agent work similarly formalised collaborative plans and commitments (Cohen & Levesque, 1991; Grosz & Kraus, 1996). These approaches show the value of making SMMs concrete and checkable, but they are often developed in controlled settings for specific tasks, known teams, and synchronous coordination. More recently, repositories, goal files, specifications, decision ledgers, and related representations are being used to expose working context to agents. In principle, repositories offer stable references, change tracking, decomposability, and machine-readable structure that can help agents interpret what is happening and why (de Halleux *et al.*, 2026). However, Gloaguen *et al.* (2026) found that context files increased inference cost and did not generally improve task success.

This suggests that representing context is necessary but not sufficient for coordinated work. Agents can respond only to what is available to them, while rationale, assumptions, and situated interpretations may remain tacit or difficult to verify. Work on common information spaces stresses that shared material becomes useful only when participants interpret it in light of their local circumstances and aims (Bannon & Bødker, 1997). Boundary objects similarly demonstrate that cooperation can occur despite different interpretations of a shared artefact (Star & Griesemer, 1989). Lee's (2007) account of boundary-negotiating artefacts further foregrounds the provisional and contested character of representations in collaborative work. Artefacts can be involved in negotiating what the work is and how it should proceed, rather than merely carrying already-established understandings across boundaries. This makes their changing interpretation part of the coordination problem. This highlights the problem of helping contributors to the work (both people and agents) (re-)interpret content as activity shifts across time, participants, artefacts, and changing circumstances.

A further challenge in human–agent workflows is that knowledge of the team must include an understanding of the capabilities of agents in the work (Andrews *et al.*, 2023), and that agents may also require computational representations of their collaborators. Explainable AI can help people form better models of a system's behaviour and errors (Bansal *et al.*, 2019) but this is often one-sided: the person learns about the AI while the AI does not maintain a model of the human. However, in collaborative settings, this awareness needs to work across the whole team: participants need shared expectations about what should be delegated, what should be reviewed, how it should be verified, and by whom. These requirements have precedent in research on joint human–agent activity, which has highlighted challenges including maintaining common ground, supporting mutual predictability, and enabling directability (Klein *et al.*, 2004).

Contemporary agents bring these longstanding coordination challenges into settings where working context can persist beyond the interactions that originally made it meaningful. Coactive Design places interdependence at the centre of human–agent system design, identifying observability, predictability, and directability as requirements for joint activity (Johnson *et al.*, 2014). On this account, effective collaboration depends not simply on participants' individual capabilities, but on their ability to understand and influence how their contributions fit together. Our concern is how this basis for interdependent action is sustained when work extends across participants, artefacts, and asynchronous episodes. Research on shared mental models, situation awareness, and joint cognitive systems has extended these concerns to human–agent teams (Jonker *et al.*, 2011; Endsley, 2023; Xu & Gao, 2023; Yao *et al.*, 2026). A process-dynamics review likewise foregrounds the co-evolution of tasks, roles, and teams across formation, development, and improvement (Wang *et al.*, 2025). Building on these accounts, we focus on how the current significance of persistent representations is established and maintained across extended, asynchronous, multi-person, multi-agent workflows.

The above suggests three requirements when considering what people and agents should understand together to act together. (i) Participants should know who knows what, who can do what, and who should be relied upon under which conditions. (ii) Coordination requires goals, plans, dependencies, and responsibilities to be clear enough to support interdependent action. (iii) Information needs to be interpreted in light of context, rationale, and assumptions. Each of these needs to be intelligible to both people and machines, and must be maintained through change.

## How can shared understanding be built and maintained as work evolves?

The previous section showed that shared understanding cannot be supported simply through access to shared representations (Bannon & Bødker, 1997). So how can shared understanding be built and maintained in collaborative work? Research into shared mental models has emphasised the role of ongoing social interaction (Cannon-Bowers *et al.*, 1993), and team cognition has been linked to the interactions through which interdependent activity is coordinated (Cooke *et al.*, 2013). Conversation is a key mechanism for building common ground, through which understanding is dynamically built and repaired (Clark & Brennan, 1991). In collaborative problem solving, the mechanisms that enable common ground have also been shown to support the progressive development of joint problem spaces, which encompass possible actions, solutions, and relevant knowledge (Roschelle & Teasley, 1995). Studies of human–intelligent virtual agent teams have reported positive associations between multimodal communication and shared mental models, and between SMMs, trust, and commitment (Hanna & Richards, 2018). Recent systems have also used generative AI to support shared mental model building among people. XSynth externalises individual models as knowledge graphs and brings them together in a shared representation; an evaluation with ten design teams reported stronger perceived SMMs alongside other design benefits (Li *et al.*, 2026).

The above indicates how shared understanding is built within episodes of interaction. CSCW research has additionally explored the question of how to maintain awareness and understanding in extended collaborative work. We have already pointed to the importance of ongoing articulation work here, which enables collaborators to manage and negotiate distributed activities, dependencies, and contingencies (Schmidt & Bannon, 1992). Formal representations of the work, such as of roles, dependencies, and plans, do not remove the need for articulation work. People interpret plans in context, responding to shifting constraints and to the practical meaning of the current situation (Suchman, 1987). The use of such representations may also be guided by conventions or protocols (Schmidt & Simone, 1996). The coordinating capacity of representations therefore depends on how people use and adapt them within cooperative work. Sustained over time, shared understanding helps people anticipate one another, coordinate implicitly, and adapt to change (Cannon-Bowers *et al.*, 1993; Klimoski & Mohammed, 1994; Mathieu *et al.*, 2000; DeChurch & Mesmer-Magnus, 2010).

Current AI agents participate in work without engaging in the rich social practices through which people establish, sustain, and repair shared understanding. Shaikh *et al.* (2024) found that, when generating conversational turns across dialogue datasets, LLMs produced fewer grounding acts than humans, instead appearing to presume common ground. Fluent contributions may therefore lack the clarification and acknowledgement through which participants establish that they understand one another. Research on grounding also identifies challenges in conveying capabilities, intentions, progress, and corrective feedback between people and agents (Bansal *et al.*, 2024). These challenges suggest that access to natural-language interaction does not itself provide the mechanisms needed to establish and repair understanding. LLMs can fail to ask clarifying questions, surface assumptions inconsistently, and repair surface linguistic ambiguity whilst missing deeper task divergence (Dombi *et al.*, 2025; Poelitz *et al.*, 2026). Likewise, people have been shown to provide infrequent and low-quality feedback to AI models, which also impacts the possibility that shared understanding can be reached (Sharma *et al.*, 2026). In human–AI interaction, interventions have been shown to improve some forms of oversight. Cognitive forcing functions, for example, can reduce overreliance in AI-assisted decision-making, although they do introduce additional interactional burden (Buçinca *et al.*, 2021).

The above emphasises that, while artefacts and records can capture goals, assumptions, roles, and dependencies, it is often communication that makes these understood, meaningful, and usable in practice. Collaborators build and sustain understanding by iteratively acknowledging, questioning, and revising information as work progresses. Interaction mechanisms that encourage critical thinking and enable feedback therefore suggest ways forward if a workable basis for action across people and agents is to be realised in practice.

## When should systems intervene to prevent or repair breakdowns in shared understanding?

Repair requires awareness of the current situation: what is happening now, what it means, and what may happen next (Endsley, 1995; Salas *et al.*, 1995). The Agent Teaming Situation Awareness framework extends this concern to bidirectional and dynamic awareness in human–AI teams, linking individual and team-level processes (Gao *et al.*, 2023). This reinforces the need to consider how people and agents contribute to one another's awareness, rather than treating transparency as information flowing only from an agent to a human observer. Maintaining this awareness is easiest in conversation, where people respond to one another in the moment (Clark & Brennan, 1991). It becomes harder in asynchronous interaction and extended collaborative work; awareness requires effort both in terms of sharing and consuming information (Dourish & Bellotti, 1992). Effective collaboration therefore relies upon the intentional, directed and often subtle communication of what really matters and to whom (Heath & Luff, 1991). This needs collaborators to understand the potential outcomes of their actions and how those interrelate with the actions of others—in other words, collaborators must consider how their actions are bound up with the dependencies (Malone & Crowston, 1994) and commitments (Bratman, 1992; Cohen & Levesque, 1991; Grosz & Kraus, 1996) that form part of the work. The importance of context is once again brought to the fore; it allows people to raise awareness in a targeted way, and to interpret when newly acquired information may suggest that they need to reconsider their own planned action.

Awareness can help collaborators share relevant information in ways that reduce incompatible courses of action, but in complex work some divergence and uncertainty may still persist. It is therefore also necessary to detect breakdowns and support recovery from them. For instance, in software development with agentic systems, four forms of oversight include a priori control, co-planning, real-time monitoring, and post-hoc review (Dhanorkar *et al.*, 2026), allowing uncertainty to be surfaced and addressed at different stages of action.

The above suggests two complementary strategies for addressing breakdown: (i) sustaining awareness through the targeted communication of relevant information to the people who need it, and (ii) building in checkpoints at key moments in a workflow, to ensure work is on track. Computational approaches have begun to infer such misalignment from task execution. Seo *et al.* (2021), for example, demonstrate Bayesian inference of team mental model misalignment in simulated healthcare scenarios, using specified task models, teammate policies, and observed actions.

### Summary

Taken together, this work suggests that shared understanding is accomplished and renewed through interaction. SMMs provide an account of relevant knowledge structures, while grounding, articulation work, and situated interpretation help explain how a workable basis for action is established, used, and rebuilt. Information that provides the relevant context for situated interpretation is likely to be distributed across personal, team, and task-specific contexts. The challenge therefore is to build understanding around what is relevant for interdependent action while preserving boundaries around information that cannot or need not be shared. The following section draws on three settings—two studies of teams and our own experience of human–agent work—to examine where shared understanding is established, how it drifts, and what enables it to be rebuilt.

## Practice Cases Informing the Framework

The conceptual framework developed in this paper draws on the related work described above alongside material from three settings: two qualitative team studies plus a reflexive longitudinal case. The cases are therefore illustrative and generative rather than confirmatory, and the framework has not yet been prospectively tested. The three settings are:

1. a scientific team experimenting with agent-enabled workflows;
2. an ML (Machine Learning) research team coordinating work through recurring planning and review cycles; and
3. the authors' own experimentation with a repository-based collaborative environment referred to by the team as the *Assembly*.

The authors have researched these three settings over the past several months to explore the use of AI in extended, collaborative, knowledge work. Two settings—the scientific work and Assembly cases—reflect intentional efforts to adopt agent-enabled workflows; the ML-research setting reflects uneven uptake of, and attitudes towards, agentic coding within the same team. In all three settings, salient findings emerged around how continuity is established, maintained, and repaired, and when intervention is required. The research was approved by the authors' organisation's ethics process. Participants gave informed consent and were not compensated.

## Scientific work: interaction changes properties that may not persist

This first study concerns a small distributed scientific team working with large-scale biological data, novel environments, and regulatory constraints. The complexity of the work has led the team to experiment with agentic workflows that rely on shared context repositories, specifications, structured question loops, and collaborative prompting.

The study followed the team's shift from ad hoc AI use to a structured, custom agent workflow, focusing on four participants in technical, domain-science, and computational roles. Data collected between March and June 2026 included two group interviews, four individual interviews, five observed collaborative working sessions, and selected Copilot chat transcripts, combining participants' accounts with records of practice. The data were analysed using AI-assisted inductive thematic analysis: AI supported initial, data-driven code generation, while the researcher reviewed, interpreted, and refined the proposed codes, returning to the raw data to check their grounding and develop and refine themes concerning participants' AI use and collaborative working practices.

### *Examples of practice*

During the period of study, the team used shared context repositories, specifications, structured question loops, and collaborative prompting to bring some consistency to their use of agents. These adaptations also supported the building of shared understanding amongst team members, including by clarifying terminology and analytical intent, but maintaining that understanding required ongoing work. Below, we describe the practices and challenges that our analysis revealed.

The repository used by the team was referred to as *Common Context*, and served as a basis for grounding, synthesis, and distributed memory. Common Context included PowerPoints, notes, specs, data definitions, device knowledge, and prior work, and enabled both people and agents to work from accumulated context. This did more than reduce cold starts: it let agents make cross-study connections and supported a rarely collocated team in maintaining awareness.

Common Context also provided an environment upon which intentional practices could be built. *Question loops* were a notable example, which encouraged explicit elicitation of information and which surfaced opportunities for repair. In a question loop, the agent would ask structured questions one at a time, requiring the team to articulate choices that might otherwise remain tacit. Especially when the team was learning together, question loops were worked through collaboratively. In these instances, the team gathered around an agent session, discussed the prompt stream, debated options, and converged on answers together. This surfaced hidden divergence, as in one example when a single question produced three different answers from three team members. The value of these sessions is therefore not just the output generated by the agent, but also the shared baseline produced by the session—team members describe leaving with a common language and a clearer shared understanding of where the work stands and which questions matter.

Question loops were one way of consolidating *specs* (specifications). Specs guided analysis and code generation. They were produced in Markdown and captured what a study is about: its research questions, constraints, relevant terminology, and decisions already made. They were executable by agents and gave them structured context, but also made relevant information legible to people in a focused way. By linking spec files to question loops, the team intentionally practised a “plan, question, question, final plan, then go on” discipline, which separated thinking from doing until intent, hypotheses, metrics, and analytical choices had been clarified.

### *Where shared understanding becomes fragile*

As discussed above, having access to shared information does not guarantee shared interpretation or even shared awareness. This was borne out here; Common Context provided a store of information, but this was not necessarily checked for updates, the repository itself did not highlight change, and agents did not reliably surface prior context at the moment it could have been relevant. So Common Context was necessary but not sufficient for the maintenance of shared understanding.

What Common Context did provide was a backdrop to collaborative work, albeit as a sociotechnical accomplishment through which the team crafted and adapted their own processes. Collaborative prompting relied on local co-prompting, shared navigation, and multi-person participation, and helped the team build awareness in shared sessions. This worked socially but was fragile. The tools involved assumed a single driver and the process could easily fall back to one person controlling the keyboard while others commented verbally. Rich alignment happened around the prompt stream—disagreements, reframings, and side decisions—but not all of this entered the prompt, spec, or agent context. Consequently, the system often failed to preserve the discussion that made the group converge.

This is one example of tacit knowledge and decision rationale being under-captured in Common Context. Session transcripts, notebooks, and shared files preserved detail, but did not draw emphasis to what mattered, what changed and why, or what was implied for next steps. The team could therefore accumulate artefacts without accumulating a usable shared account of the reasoning behind them.

Particularly salient examples of this within our data centred on collaborative handoffs. Handoffs were especially important for the team, as their differentiated expertise across biology, statistics, data infrastructure, and ML engineering meant that specific people needed to play particular roles. For instance, verification often required specific expertise, which meant bringing different team members in at different moments in the work. However, sharing agent-generated code and analyses often communicated the “what” of the work but not necessarily the “why” needed for another person to continue it. Effective handoffs need to make uncertainty and next steps clear, and must translate this across disciplines. In this team, translation was improvised by individuals, sometimes resulting in lost intent, incomplete context, or insufficient emphasis being drawn to relevant prior knowledge.

### *Repair: attempted, absent, and needed*

Repair was driven through social practice as well as infrastructure: question loops surfaced divergence and collaborative sessions rebuilt a shared baseline. However, handoffs where translation across disciplines was needed presented challenges, especially where the discussion that made the group converge did not enter the agent context.

This case illustrates that, in collaborative work, shared understanding built through social interaction can remain separate from the context provided to agents, and agent contributions can introduce uncertainty into handoffs across team members. Deliberate infrastructure can support shared understanding, but only when coupled with intentional social practice adopted at a team level. Further changes to social practice could alleviate some of the problems highlighted: social interactions could be manually transferred to shared records; handoff moments could be collectively redesigned so as to help reconstruct significant context across participants. Agents themselves could also be adapted so as to better leverage knowledge of the team.

## ML research: coordination properties drift between synchronisation points

The second study explored the collaborative working practices of a multidisciplinary ML research team, with the aim of understanding how members coordinate and maintain shared awareness, and where AI could support more effective collaboration. Seven team members in research and management roles took part in the study, which ran from February to June 2026. Initial data collection included a workshop with the team and seven semi-structured individual interviews, combining collective and individual perspectives on the team's working practices. The data were analysed collaboratively by a researcher and Copilot: both examined the data independently to identify themes and UX opportunity areas, and the researcher then compared, challenged and combined these perspectives, returning to the raw data to check their grounding. Finally, a follow-up survey invited the team to validate and prioritise the opportunity areas, ensuring the analysis reflected participants' own views of their needs.

### *Examples of practice*

For this team, shared understanding is maintained largely through meeting cycles. They organised their work over quarterly objectives, two-week sprints, fortnightly sprint reviews, and weekly feature-team meetings. These recurring meetings provided a rhythm through which retrospection and prospection connected otherwise distributed threads of work (Vanukuru *et al.*, 2025). Quarterly planning aligns goals and strategy; sprint planning establishes shorter-term commitments; feature-team meetings coordinate local work; and reviews reconstruct what happened. These touchpoints resulted in specific artefacts. Quarterly and sprint review presentations accumulated over time, preserving a historical record of decision outcomes, priorities, and completed work. Written OKRs served as goal anchors, taking the form of explicit shared objectives meant to guide sprint work and provide a reference point for local prioritisation decisions throughout the quarter. The study focuses on the team's own accounts of episodic synchronisation and friction as shared understanding was periodically reconstructed after intervals of distributed work, exploring both what these touchpoints preserved and what remained fragile between them.

### *Where shared understanding becomes fragile*

The meetings were necessary because the team's distributed work often caused interpretations to diverge. Team members described situations in which everyone believed they were pursuing the same objective while holding subtly different interpretations of priorities, success criteria, or next steps.

This challenge was exacerbated by knowledge that was fragmented and implicit. Relevant context was distributed across Teams channels, PowerPoint decks, GitHub repositories, and personal notes. No single source of truth was actively maintained between meetings, so the integration of distributed knowledge was always partial and delayed. Furthermore, why something was decided—the trade-offs considered and the alternatives rejected—was rarely explicitly captured from meetings, instead living in the memory of attendees or being lost through forgetting or turnover. If team members missed a meeting, shared context needed to be reconstructed, which was often accomplished through conversation and so also produced no durable shared record.

Thus, as work progressed, divergence could accumulate while often going unnoticed until the next sync. Team members described how this could be amplified when agents were used to generate code and analyses. Agent use was reported to increase output volume, widening the gap between what was produced and what any one individual understood. There was no mechanism for agents to cross-check outputs against a shared picture of intent.

The study foregrounds cycles of alignment, drift, and reconstruction of shared understanding. Participants often believed they were working towards the same goals while operating from subtly different interpretations, leading to divergence without explicit disagreement. Shared understanding was therefore dependent not only on access to information, but on preserving the rationale and meaning that connected decisions, plans, and artefacts. Meetings emerged as an important site of collective sensemaking, where shared interpretations were revisited, negotiated and rebuilt rather than simply being communicated. This periodic synchronisation can be understood partly as work through which participants reconstructed the *current* significance of persistent representations for coordination.

### *Repair: attempted, absent, and needed*

In the ML research team, repair happened largely at synchronisation points. Between meetings, repair was largely absent. The team's own analysis suggested that episodic alignment became insufficient as information volume was growing, making it difficult to manually track work. This suggests a need for repair that is sustained between meetings, rather than periodic reconstruction.

## The Assembly: representations persist while their properties change

The third example is taken from the authors' own working environment, *The Assembly*, an evolving, repository-based environment that brings together shared goals, research materials, meeting records, agent instructions, and custom tools to support collaborative knowledge work between people and AI agents.

We used the Assembly between March and October 2026, and captured data about our usage via agent-facilitated individual reflections, workflow-specific reflective notes, and transcripts of working meetings and collective discussions. These exist alongside a persistent and version-controlled record of goals, specifications, agents and agent instructions, generated artefacts, and context such as transcripts of every meeting conducted during the period of study. Workflows focused on preparing presentations, writing proposals and position papers, and conducting analyses of the two studies described prior to this one. The team also adopted specific proposals to articulate and capture individual and team goals, and developed a number of agents and prototypes to support their work. Together, these sources provide a basis for a reflexive longitudinal account, relating participants' descriptions of benefits, difficulties, and changing practices to the artefacts and discussions through which their work developed.

### *Examples of practice*

The team stores goals, specifications, meeting transcripts, key documents, and agent-generated artefacts in the Assembly. We do this intentionally to create a shared context that encompasses as much of our work as possible in a form that both people and agents can consult and act upon. Markdown makes goals machine-readable, specs translate intent into concrete expectations, bespoke agents work with goals, specs and the context provided by the Assembly to generate content, and tools built by team members are used to render that content in forms that can be edited, presented, and interacted with. For example, in preparing to show their work over a series of meetings, the team created goal and spec files to ground the generation of content by a tailor-made agent. The agent's contributions could be displayed using a custom-made piece of software, which enabled the team to iterate upon the presentations themselves as well as deliver them to others.

The Assembly's infrastructure was designed to make the team's intent machine-queryable as well as accessible to people. Agents plan their work while referring to goals that have been recorded by the team, utilising specs that also help them align on conventions and constraints that should shape how work is carried out. While some workflows could be described as 'one-shot', with agents drawing on goals, specs, and context to generate an artefact, team members also experimented with workflows that brought in agents to support iteration throughout the longer processes of working on papers. In these cases, agents were tailored to wrap their additions in explicit markers and link them to ledger items, making AI contributions distinguishable from human-authored material. This made it easier to develop content more slowly over time, working with agents as well as the wider team, whilst reducing the risk that generated content would silently enter the Assembly without attribution or review.

### *Where shared understanding becomes fragile*

While the team made a concerted effort to capture the context of their work and make it readable by agents, not everything that was captured remained current. The team's social conventions to articulate goals as part of their work required ongoing social maintenance, and this needed to be performed in an environment that was undergoing constant change. New agents and tools to enable the expression and discussion of goals were introduced by different team members over the period of study, some being used by individuals in support of their own practice, whilst others were made available to the team. However, having all team members align on ways of articulating goals was an ongoing challenge.

Furthermore, goals and specs that guided work often remained static, while the reasoning around the work emerged and evolved through its progression. The intent represented in the repository could therefore be stale, fragmented, and even plural. Agents referred to goals and specs first, goals being first-class citizens in the Assembly. For team members, insights could form and intent could be developed both when working with artefacts—including in iteration with people and with agents—as well as through discussion during meetings. However, while meeting transcripts formed part of the Assembly's context, there were many of these (186 at the time of writing), and records of human–agent interaction did not enter the wider Assembly. The volume of content in the Assembly, coupled with the opacity of knowing what is 'in' context, made it difficult to determine the common ground for action when working with agents.

### *Repair: attempted, absent, and needed*

In the Assembly, repair relied on explicit markers and social conventions. Perhaps most salient was the absence of repair of stale intent. The study demonstrates the value of persistent representations, but also their dependence on continuing articulation work. Goals, specifications, ledgers, artefacts, and commits were sites where shared understanding was made visible, maintained, but also sometimes lost. Externalising shared state did not remove the need to maintain its current meaning through ongoing work and interaction, or to extend this maintenance work to agents' interpretations of what is relevant now.

Taken together, the three cases show different ways in which continuity can fail. In the scientific team, understanding built through question loops and collaborative prompting did not reliably reach the agent context or survive handoffs, so the "why" behind work was often left behind. In the ML research team, shared understanding was rebuilt periodically at meetings, but interpretations drifted between synchronisation points, and agents widened the gap between what was produced and what any individual understood. In the Assembly, representations persisted and remained machine-readable while the intent around them evolved, so that goals and specifications could become stale, fragmented, or plural. In each case, much of the reasoning behind decisions, and much of what was learned through discussion, was not captured where other people, or agents, could find it. Repair, where it occurred, depended on social practice at particular moments (sessions, meetings, markers), and was largely absent between them. In none of the cases was access to shared information enough.

# A Framework for Coordination Continuity

Building on related work and the three cases described above, we highlight the need for contributors to build, maintain, and update mental models of task and team, and to articulate, test, and repair understanding that is relevant to current action. In this section we present a framework that distinguishes representations of work, their properties for coordination, and the interaction mechanisms through which those properties become visible and can be changed. The framework assumes that coordination continuity is a sociotechnical accomplishment; information and context relevant to a collaborative effort may be distributed across people, agents, artefacts, repositories, histories, and local circumstances. No single participant or representation needs a complete model of the work, but they must understand how their own actions intersect with that wider context.

## Representations of collaborative work

Representations are relatively durable objects around which collaborative activity can be organised. Extending SMM theory and building on our own analyses, we suggest that key representations for human–agent teamwork should include:

### *Goals and intent*

*What is the collaborative effort trying to accomplish, and what constitutes an acceptable outcome?*

Goals provide reference points against which people and agents can interpret local actions. Making goals explicit and machine-readable can help agents orient their activity (Molinaro *et al.*, 2026). However, representing a goal does not ensure that participants interpret it compatibly or that the representation remains current over time; this requires ongoing work.

### *Plans, commitments, and dependencies*

*What is intended to happen, who is doing what, and how are actions interlinked?*

Plans represent intentions for how activities, actors, and resources relate to one another. They signify a form of joint commitment between participants (Cohen & Levesque, 1991; Grosz & Kraus, 1996) and help them understand how local actions relate to the broader work (Suchman, 1987). This then provides a foundation for ongoing articulation work and coordination (Schmidt & Bannon, 1992).

### *Roles, capabilities, and authority*

*Who or what knows what, can do what, and can make decisions?*

Effective collaboration can depend on differentiated expertise rather than knowledge overlap (Wegner, 1987; Hollingshead, 1998; Lewis, 2003), and so coordination also depends on understanding who or what knows what, can do what, and has the authority to make or confirm particular decisions. In human–agent teamwork, this extends to knowing an agent's capabilities and the requirements for verification that its work introduces. As demonstrated by the case study of the science team, expertise that is differentiated across a team may require a single agent output to be verified by more than one team member.

## Coordination properties

Representations need to be interpreted in relation to the current state of work and what should happen next. This requires understanding what has happened, what remains open, and where things currently stand. We propose that this understanding is reached by considering the *coordination properties* of a representation: aspects of its status, such as whether a goal is current or a plan is committed, that bear on how it should be used now.

Coordination properties are:

- Relevant to interdependent action; they matter because one participant's action depends on, or affects, another's. For example, whether evidence has been verified or not is important if another participant intends to rely on it.
- Relative; they may have a different status across participants. For instance, one participant may view evidence as sufficiently verified while another does not. Visibility of difference is important in collaborative work.
- Dynamic; they are not fixed labels but may be revisited as actions change the state or meaning of the collaborative effort. For example, a goal that was current when an agent acted on it may later be superseded.

In this section, we provide illustrative coordination properties for each of the representations described above (goal, plan, and role), to show the utility of coordination properties but also to draw out their contingency on who is working with a representation, and when. We use as a unifying example the development of materials to support a meeting with an engineering stakeholder by the team working in the Assembly.

For this meeting, the team agreed that the aim would be to demonstrate their technology and explore possibilities for collaboration. This *goal* was represented in a Markdown file that explicitly documented the meeting purpose. The *plan* was likewise written in Markdown (as a spec), which was generated through interaction with an agent that also had access to the goal file. The spec set out an aspirational meeting outline and timeline. The *role* we explore here is that of a presentation planning agent, which was configured to generate presentations from goals, specs, and named sources. The agent was created to support a number of meetings that were scheduled over the course of one week and that would each need its own preparation work.

Use of these representations was driven by people at specific moments in the workflow, and knowledge of this timeline is important when interpreting their status. For instance, when used to generate the spec, the goal file could be said to be *current*. The same is true for the spec when it was used to generate a presentation. At these moments, the representations were judged to provide a sufficient basis for work to progress; they guided action and narrowed alternatives. It is worth mentioning however, that even at these moments, goals and specs might have differing degrees of *acceptance* or *opposition* across team members; the audience for the presentation might also have different intentions and plans for the meeting. So even if a person responsible for the work drives it forwards, there may exist variation in how widely those representations are endorsed.

Content was often iterated following its generation. A spec could be developed further following its generation from a goal file; a presentation could be improved following its generation from a spec. The degree of iteration, and the resulting relationships between the goal, spec, agent and other sources, may indicate whether the goal and/or plan were *upheld* through iteration, or became *superseded* as the work developed. Later reflections on how the presentation was received may also indicate how the meeting went, and so whether the goal and plan were upheld in practice. The same relationships might indicate whether the agent was *fit for its role* or whether it would be *in need of revision* if it were to be reused. For goals and specs for particular events, such as meetings, there is a likelihood that they become *historical* shortly afterwards; this too has implications for how they might be reused.

The example outlined here underlines how coordination properties are linked to *interdependent action*. Within this human-driven workflow, a person took responsibility for deciding when a goal or plan was sufficient for an agent to act upon it. This suggests that, *at that moment*, the representations involved had a certain status. However, if any of these representations were to be acted upon later, that interpretation would need to be made afresh, and would likely need to take account of *when* the representations were acted upon, by *whom*, for *what* purpose, and *what had happened since*.

Drawing from the above, example coordination properties include:

- Current or historical: whether a representation is a sound basis for action now. The goal and spec were current when used to generate an artefact, and were likely to become historical shortly after the meeting they were associated with.
- Upheld or superseded: whether a representation held up through iteration. The goal and spec could be upheld as the work developed, or become superseded by changes to the outputs they generated.
- Accepted or opposed: how far a representation is endorsed. Team members may show differing degrees of acceptance or opposition, and the meeting audience may have had different intentions.
- Fit for role: whether a participant (human or agent) is performing the role they have been given, or whether the role needs revisiting. The presentation planning agent, for instance, might be fit for its role for the meeting it was configured for, but in need of revision once relevant sources have changed.

These examples are illustrative rather than exhaustive. Other properties matter in other circumstances, such as whether evidence has been verified, or whether a representation is considered authoritative enough for an agent to act on proactively. Whichever properties are in play, interpreting them depends on provenance: participants need to understand not only what information is available, but how it is evidenced. Provenance can support interpretation by showing why information was considered relevant, why one interpretation was chosen over alternatives, and what was rejected. Surfacing this status, and changing it, is the work of interaction mechanisms, which we turn to next.

## Interaction mechanisms

Interaction mechanisms are the processes through which participants establish, update, and rebuild their basis for coordinated action. They enable people and agents to interpret whether to proceed, seek clarification, or revisit how things now stand with the work before they act.

Interaction mechanisms play two related roles. First, they can foreground coordination properties that would otherwise remain latent, raising awareness of uncertainty. For example, it might be important to highlight that two participants interpret a goal differently, or that a plan has been superseded. Second, interaction can change representations and their properties, as when evidence moves from unverified to verified, through attempts to (re-)build the understanding needed to enable further action.

Examples include the following, of which the first two foreground coordination properties and the rest help to change them:

- Mechanisms that surface missing evidence, expertise, or alternative interpretations before participants proceed;
- Mechanisms that make the source material an agent is drawing on visible, so that differing interpretations of it, over time or across documents, can be noticed;
- Mechanisms that help people or agents seek clarification and verification;
- Mechanisms that help people provide informative feedback that highlights and corrects misunderstandings (cf. Sharma *et al.*, 2026);
- Mechanisms that carry the rationale and outcomes of interaction into the shared record, so that those who were not present, including agents, can interpret them.

Because awareness carries costs both for those who share information and for those who attend to it (Dourish & Bellotti, 1992), mechanisms that foreground coordination properties need to be selective about what they surface, and to whom. Mechanisms including clarification, verification, negotiation, and escalation, have been widely researched in relation to human grounding and articulation work (Clark & Brennan, 1991; Schmidt & Bannon, 1992). There is a need to tie this work to human–AI interaction, and to apply it to settings where people and agents alike are working with particular representations as grounding for the work. These mechanisms are especially relevant when a basis for action is built and rebuilt, as the next section explores.

## Build → Act → Drift → Rebuild

The above shows how representations, coordination properties, and interaction mechanisms connect to support coordination continuity. Here, we highlight *build → act → drift → rebuild* as a heuristic that underscores why continuity must be maintained over time.

- Build: Participants establish a sufficient basis for interdependent action. This may involve articulating intent, identifying expertise and dependencies, and documenting disagreement. Complete consensus is not necessary, however participants must understand how differences in interpretation affect the next action. An example is provided by the science team's question loops, which required members to articulate choices that might otherwise have remained tacit.
- Act: Participants act on the shared basis that has been established. This action may change representations, their properties, commitments and dependencies. In the Assembly, using a goal to generate a spec, and the spec to generate a presentation, committed the team to each representation at that moment, but subsequent iteration could leave them upheld or superseded.

- Drift: Differences emerge between participants' interpretations of the work. This may be productive; however, drift becomes problematic when differences imply incompatibility. Judging when this threshold has been crossed is not straightforward. In the ML research team, members believed they were pursuing the same objective while holding subtly different interpretations, and divergence often went unnoticed until the next sync.
- Rebuild: Participants renew a workable basis for action. This may be necessary even in cases of productive difference. Rebuilding may preserve multiple interpretations. What matters is that participants leave with enough common ground to take the next step. Recurring meetings in the ML research team played this role.

The stages in this heuristic need not be linear; grounding and repair moves occur throughout collaborative work. However, we do wish to foreground a temporal relationship here. Persistent representations can outlive the circumstances that once made them relevant, while action continually creates new circumstances under which that relevance must be reinterpreted. Continuity is most at risk at what we call *consequential moments*: points at which action rests on a representation whose current status is uncertain, and where acting on a mistaken interpretation would be costly to undo or would spread to others. Kowalyshyn and Scheutz (2026) distinguish unsupported beliefs, false beliefs, belief contradictions, and omissions, and examine their patterns across repeated dyadic tasks. Such distinctions could inform the detection of candidate breakdowns, while leaving open a further judgement: whether a discrepancy matters for the interdependent action about to occur. This connects to the two strategies identified in related work. Awareness helps participants notice drift as it happens, while checkpoints place opportunities for rebuilding where drift would matter most.

### Summary

Taken together, representations, coordination properties, and interaction mechanisms give participants a way of judging whether rebuilding is needed before action proceeds. Collaborators can ask whether a representation, or their understanding of its status, has drifted; whether upcoming action makes that difference consequential; and, if so, how the difference can be surfaced, tested, or negotiated so that the next action rests on a workable basis. The design implications that follow consider how collaborative AI systems might support this judgement and the interaction it requires.

## From Framework to Design

We began this paper by asking what people and agents need to understand together to act together, how that understanding can be built and maintained as work evolves, and when systems should intervene to prevent or repair breakdowns. Building on prior work, two qualitative team studies and one reflexive case, we have developed a framework that distinguishes representations of collaborative work, the coordination properties that support interpretation of those representations, and the interaction mechanisms through which representations and properties are surfaced and changed, together with a *build → act → drift → rebuild* heuristic to illustrate how and why coordination continuity must be maintained over time. In this section, we translate the framework into design implications for collaborative AI infrastructure: one that supports interpretation rather than retrieval, while preserving differentiated expertise and productive difference. This draws on what CSCW has long shown: shared artefacts do not by themselves sustain collaboration, because they are continually interpreted and articulated (Bannon & Bødker, 1997; Schmidt & Bannon, 1992) and act as resources for situated action (Suchman, 1987). As agents act on persistent goals, specifications, and plans without the social practices through which people question and repair shared understanding, the articulation work that people routinely do can no longer be taken for granted.

Agent infrastructure should therefore complement, rather than replace, the interpretive work that people do. Repositories, tools, and feedback loops should be connected to the interaction mechanisms through which people and agents establish congruent understandings, so that systems support articulation and judgement rather than invisibly resolving differences on participants' behalf.

The eight implications that follow build on the framework to suggest a path for design, organised around the three questions that frame the paper. The first three concern what must be understood together: supporting interpretation rather than retrieval, surfacing roles and expertise, and representing uncertainty whilst preserving productive difference. The next three concern how that understanding is built and maintained: capturing rationale and the outcomes of social interaction, supporting clarification and feedback, and maintaining the conventions through which representations are interpreted. The final two concern when systems should intervene: making coordination status selectively visible, and triggering proportionate rebuilding before consequential action.

## 1.1 Support interpretation, not simply retrieval

Collaborative AI systems should help participants understand the current status of the representations their work relies on. Related work suggests that representing context is necessary but not sufficient for collaboration, and the Assembly case illustrates this: its repositories held material that was no longer current. A system might help a participant—human or agent—ask: Is this goal still current? What changed since this decision was recorded? What evidence supports this position? A representation's status may be interpreted from interaction histories, artefacts, and knowledge of team conventions. In human–AI interaction, this also means making visible what is in an agent's context window, so that the common ground for action is clear. Systems should likewise surface provenance and make clear the inferential assumptions connecting AI-generated claims to their sources (Sarkar *et al.*, 2026). Taken together, these capabilities shift the system's role from retrieving what was recorded to helping people and agents judge whether what was recorded still holds for the action at hand.

## 1.2 Make roles, expertise, and verification responsibility legible

Effective collaboration depends on knowing who or what knows what, can do what, and has the authority to decide (Wegner, 1987; Hollingshead, 1998; Lewis, 2003). In the science team, differentiated expertise meant that a single agent-generated analysis could require verification by multiple people, yet shared code and analyses rarely made clear who needed to verify what, or why. There are two broad ways to make capabilities intelligible. The first is to represent them explicitly, for example as a description of what an agent or team member can do. This makes capabilities easy to inspect and to route work against, but risks becoming stale. The second is to let capabilities be learned and negotiated through interaction. He *et al.* (2023) built a working system that gives people and AI a shared picture of what each can do, then uses repeated task grouping and negotiation to decide who takes which task. This keeps the picture grounded in how participants actually perform, although it can leave that knowledge tacit and hard for others, including later participants, to see. Systems should make roles, capabilities, and verification responsibilities visible, and help route requests for clarification or verification to the participants best placed to answer them. Because whether a participant, human or agent, is fit for its role is itself a coordination property, systems should also help teams notice when a role needs revisiting as sources, tools, or the work change.

### 1.3 Represent uncertainty and preserve productive difference

Coordination properties should not be treated as objective labels automatically attached to artefacts. A system may encounter, for instance, conflicting assessments of whether evidence is verified, uncertainty about whether a goal remains current, and incomplete evidence about whether a decision is committed. Systems should surface uncertainty and help participants determine what is “current”, “verified”, or “accepted”, while recognising that this may privilege some interpretations over others. Disagreement does not always need to be resolved. Teams benefit from independent perspectives and specialised expertise (Wegner, 1987; Hollingshead, 1998), and the ML research team’s experience shows that harmful misalignment can arise without any explicit disagreement at all. Systems should therefore distinguish between productive difference, where participants know or believe different things but can still coordinate, and harmful misalignment, where differences imply incompatible expectations for interdependent action, and should preserve alternatives rather than silently collapsing them.

The second group concerns *how* shared understanding is built and maintained as work evolves: by capturing the reasoning behind decisions, through clarification and feedback, and through the conventions that sustain these practices.

### 2.1 Capture rationale and the outcomes of social interaction, selectively

Across all three cases, the reasoning behind work was under-captured. In the science team, the alignment that made the group converge happened around but outside of the prompt stream. In the ML research team, why something was decided was rarely documented from meetings. In the Assembly, records of human–agent interaction did not enter the wider Assembly, and insights developed through iteration and discussion were not reflected in static goals and specs. Systems should help teams capture not only decisions but the reasons for them (see Park & Lee, 2023, for a similar point in relation to communicating the reasoning behind edits in collaborative writing). For human–agent workflows, this suggests connecting a change to the reasons behind it, so that later contributors can examine the interpretation behind the result. Design-rationale research explicitly considers the costs as well as the uses of representing arguments and decisions (Buckingham Shum & Hammond, 1994). The aim should therefore not be to record every exchange, but to preserve reasoning whose later absence would make a decision difficult to assess, and making that inspectable and revisable.

### 2.2 Support clarification and feedback as part of the work

Agents can fail to participate in conversational grounding (Dombi *et al.*, 2025; Poelitz *et al.*, 2026), and people are poor at giving feedback in human–AI interaction (Sharma *et al.*, 2026). The science team’s question loops show what clarification can look like when it is built into a workflow, foregrounding interpretations and differences that might otherwise have remained tacit. Systems should therefore treat clarification as a first-class interaction. Agents should ask about goals, assumptions, and dependencies where these are ambiguous or contested, and surface the assumptions on which they are acting. Systems should also help people give feedback that is informative. Clarification has costs, and interventions that reduce overreliance can add interactional burden (Buçinca *et al.*, 2021), so questions should be targeted at points where the answer would change what happens next, a judgement that connects to consequential moments (see below).

## 2.3 Support the maintenance of collaborative conventions

The Assembly case showed that representations remain useful only while participants maintain the conventions through which they are interpreted, and that these conventions need continual social maintenance as tools and agents change. Systems might therefore help teams identify when conventions are no longer being followed, and support their revision as work, tools, and team composition change. This includes conventions for recording rationale, marking and verifying AI contributions, and updating goals and specifications. Agents themselves may participate in this maintenance work, rather than assuming that stored representations remain current and meaningful.

The third group concerns *when* systems should intervene: by sustaining awareness between moments of synchronisation, and by prompting rebuilding where action becomes consequential.

## 3.1 Make coordination status selectively visible

Drift between synchronisation points, as in the ML research team, suggests that awareness needs to be sustained between meetings rather than only at them. However, more awareness is not necessarily better, as it takes effort both to raise awareness and to attend to information (Dourish & Bellotti, 1992). Earlier human–agent research has explored this selectivity through cognitive-load-sensitive information sharing and shared belief maps (Fan & Yen, 2011). It illustrates how supporting an evolving SMM involves decisions about what to communicate under particular conditions, rather than simply making all information available. Systems should surface coordination properties according to their relevance to the participant, the purpose for which they are needed, and the risk involved. The design problem is to make enough relevant coordination status available for the interdependent action at hand.

## 3.2 Prompt proportionate rebuilding at consequential moments

As the *build* → *act* → *drift* → *rebuild* heuristic suggests, not every divergence needs repair, but some actions are consequential enough to make rebuilding a requisite for subsequent action. Systems should weigh whether additional alignment work is needed, using signals such as limited reversibility, transfer of responsibility, and commitment to one interpretation. At these moments, systems should invite human attention. Handoff and re-entry are particularly clear consequential moments, because of the need to reconstruct enough collaborative state to progress the work. Systems should support reconstruction not only of what changed but of how relevant representations might now be interpreted. In the science team, handoffs often communicated the "what" of agent-generated work but not the "why". A successful handoff restores a workable basis for coordinated action rather than merely transferring information.

## Limitations

The practice cases described here are drawn from small qualitative studies and from the authors' own working environment, which also brings risks of familiarity. The cases informed the framework but were not designed to test it, and the framework has not been applied prospectively. Analysis in the two team studies was AI-assisted, with researchers reviewing proposed codes and themes against the raw data, and the Assembly account draws partly on agent-facilitated reflections; these methods may shape which themes are surfaced. All three settings involve knowledge work with software-based agents, so the findings may not transfer to other domains, team sizes, or agent capabilities. A review of 88 empirical studies shows substantial diversity in how mental models of AI are defined and elicited, and treats elicitation itself as an interpretative process that may influence what is elicited (Sanchez *et al.*, 2026). Our interviews, reflections, and work records should accordingly be understood as situated evidence about coordination, rather than direct measures of participants' internal models.

The coordination properties we describe are illustrative rather than exhaustive, and the signals we propose for consequential moments are candidates, drawn from related work and our cases. Research is needed to investigate which representations and properties matter in different forms of collaborative work. A scoping review of users' mental models in human–AI interaction likewise identifies conceptual and methodological diversity and calls for greater attention to context and longitudinal investigation (Bodamer *et al.*, 2026). Evaluating coordination continuity will require following how interpretations and dependencies change across episodes, rather than assessing shared understanding only at a single point in time.

## Conclusion

Extended human–AI collaboration increasingly depends on persistent representations: goals, plans, specifications, and records that both people and agents can read and act upon. Making these explicit is valuable, but it does not by itself keep a team coordinated. The work around a representation moves on while the representation itself endures, and its significance can change without anyone, or anything, noticing.

We have proposed coordination continuity as a way of framing this problem: the extent to which a workable basis for interdependent action is sustained as work evolves. Drawing on prior research and on three practice cases, we present a framework of three elements: representations of collaborative work; the coordination properties that indicate how those representations should currently be interpreted; and the interaction mechanisms through which those properties are surfaced and changed. We added a *build → act → drift → rebuild* heuristic, and the notion of consequential moments, to show why this work is continuous and where it matters most.

Returning to our three questions, we have highlighted that people and agents need to understand together who is doing what and who can be relied upon, how goals, plans, and dependencies currently stand, and how information should be interpreted in light of its context and rationale. This understanding needs to be congruent, not identical, with differences made visible rather than hidden. It is built and maintained through interaction mechanisms such as clarification, verification, feedback, together with the capture of reasoning and understanding of social conventions, and it must be rebuilt as action changes the work. Systems should intervene proportionately, by sustaining awareness between moments of synchronisation and by prompting rebuilding at consequential moments such as handoff and re-entry. The eight design implications translate these answers into directions for collaborative AI infrastructure.

Persistent memory can be built, but persistent understanding needs to be maintained. The aim for human–AI teams is not to remember more, but to notice sooner when what is remembered no longer fits the work.